\documentstyle[12pt]{article}

\begin{document}

\begin{flushright}
NUB 3102\\
October 1994
\end{flushright}
\pagestyle{empty} \vspace{1in}
\begin{center}
{\Large {\bf Another Look at the Einstein-Maxwell Equations}}\\
\vspace{0.5in} {\sc M. H. Friedman}\\ \vspace{0.3in} {\it Physics Department
\\ Northeastern University\\ Boston, MA 02115, USA\\ \vspace{1in}ABSTRACT}\\
\bigskip
\parbox{5.2in}
{An electric monopole solution to the equations of Maxwell and
Einstein's general relativity is displayed. It differs from the usual one
in that all components of the metric vanish at large spatial distances
from the charge rather than approaching the Minkowski metric. This enables
us to find an approximate solution to that for many charges.}
\end{center}

\newpage
\pagestyle{plain} \pagenumbering{arabic}

\section{\bf Introduction.}

A great deal of work has been done investigating the properties and
symmetries \cite{1} of the Einstein Maxwell system of equations. However,
few solutions have been found to date \cite{2}. In this paper I present a
very simple static, exterior solution, in which the components of the metric
vanish asymptotically. The field is that of a spherically
symmetric electric charge distribution.

Such a solution is of interest as it is describes a universe consisting of
an electric charge and a metric for which all space time points are light
like separated at large distances from the charge. The origin of this
property then permits us to find an approximate solution to the case
involving many charged particles.

The solution was found with the help of an algebraic computer program.

\section{\bf The Solution.}

The equations to be solved are \cite{3}
$$
R_{\mu \nu }-\frac 12g_{\mu \nu }R=8\pi T_{\mu \nu }\eqno (1)
$$
$$
F^{\mu \nu };\nu =4\pi J^\mu \eqno (2)
$$
where

$$
T_{\mu \nu }=\frac 1{4\pi }\left[ g^{\alpha \beta }F_{\mu \alpha }F_{\nu
\beta }-\frac 14g_{\mu \nu }F_{\alpha \beta }F^{\alpha \beta }\right] ,\eqno
(3)
$$
$R_{\mu \nu }$ is the contracted curvature (Ricci) tensor, $R$ the scalar
curvature, $g_{\mu \nu }$ the metric tensor, $F_{\mu \nu }$ the
electromagnetic field strength, and $T_{\mu \nu }$ the Maxwell stress-energy
tensor.

Eq.(2) may be written as
$$
\frac 1{\sqrt{-g}}(\sqrt{-g}F^{\mu \nu }),_\nu =4\pi J^\mu \eqno (4)
$$
where $g$ is the determinant of the metric tensor.

The metric is conformally related to that of flat space so that it is given
by the line element
$$
ds^2=h^2(dr^2+r^2d\theta ^2+r^2\sin ^2\theta d\phi ^2-dt^2).\eqno (5)
$$

The field strength is given by%
$$
F_{\mu \nu }=A_{\nu ,\mu }-A_{\mu ,\nu }\eqno (6)
$$
where%
$$
A_0=\frac er\eqno (7)
$$
for $r>r_0$ and $A_i=0$ for $i=1,2,3$.

A static monopole solution to these equations for $r>r_0$ is%
$$
h=\frac dr\eqno (8)
$$
with
$$
d=\frac{e\sqrt{G}}{c^2}\eqno (9)
$$
where $d$ and $e$ are constants, and $G$ and $c$
have been temporarily restored for the purposes of clarity \cite{4}.

$J^\mu $ is that of a static, spherically symmetric, electric charge
distribution contained inside the region $r\leq r_0$ and of total charge $e$
. This can be verified by assuming the existence of an interior solution
that can be joined to this exterior one. One then integrates Eq.(4) over a
spatial volume enclosing the charge, using appropriate factors of $\sqrt{-g}$
, {\it i.e.},

$$
q=\int_{x^0=const.}\sqrt{-g}\times \frac 1{4\pi \sqrt{-g}}(\sqrt{-g}
F^{0i})_{,i}\,drd\theta d\phi .\eqno (10)
$$

Since, the integrand has only a $\sin\theta$ dependence on angle, the
integral reduces to the value of $\sqrt{-g} F^{0i}/\sin\theta$ evaluated
at $r$ ($r>r_0$), whereupon Eqs.(5) and (7) are then used to find $q=e$.
(I have assumed that $\sqrt{-g} F^{0i}/\sin\theta=0$ at $r=0$ which will
be true except for the point charge where it
equals $e$. However, the point charge is not considered in this paper
since it is not a solution to the integrated form of Eq.(1).)

It is interesting to note that an electrically neutral test particle at rest
in this universe will have an energy $E=m\sqrt{-g_{00}}=mh$. Hence $E$ will
vanish at large spatial coordinate distances from the charge.

\section{Comparison to the Reissner-Nordstr\"om Solution.}

The Reissner-Nordstr\"om solution \cite{5} is the spherically symmetric,
static, exterior field of a charged distribution of mass. It is given by

$$
ds^2=\frac{dr^2}l+r^2(d\theta ^2+\sin ^2\theta d\phi ^2)-ldt^2\eqno (11)
$$
where

$$
l=1-\frac{2M}r+\frac{e^2}{4\pi r^2}.\eqno (12)
$$

If $M$ is set to zero we can then make a direct comparison between Eqs.(5)
and (11). In Eq.(5) all the components of the metric vanish for large
distances from the charge, so that $ds^2\rightarrow 0$ while in Eq.(11)
$ds^2 $ approaches the Minkowski expression for flat space.

On the other hand as $r\rightarrow r_0$ and $r_0\rightarrow 0$, Eq.(5) gives
$ds^2\rightarrow (d/r)^2[dr^2-dt^2]$ which can be light-like, space-like or
time-like while Eq.(11) gives $ds^2\rightarrow -(e^2/4\pi r^2)dt^2$.

We also note that a test particle at rest in the universe of the metric of
Eq.(11) will have an energy $E=m\sqrt{1+e^2/4\pi r^2.}$and hence approaches
$m$ at large distances from the charge in contrast to the energy of the test
particle in section 2 which vanished at large distances. However, the
energies have the same behavior at very small distances.

\section{An Approximate Solution for Many Charges}

Referring to Eq.(5) we note that $ds^2$ goes like $d^2/r^2$. Let us now
consider charges of the order of magnitude of that of the electron.
Referring to the values given in Ref.[4], we see that if we are at atomic
distances, {\it e.g.}, the Bohr radius, from the charge, $ds^2$ will have
dropped by a factor of $d^2/r^2\approx 10^{-52}$ from what it was in the
neighborhood of the charge. Thus if we have a gas of these charges,
separated by atomic distances, the metric will either be effectively zero if
we are away from any charges, or will be given by Eq.(5) (with the origin
shifted to the location of the $i^{th}$ charge) if we are within a distance
$d$ of that charge. Hence, in this approximation%
$$
ds^2=\sum_ih_i{}^2(dx^2+dy^2+dz^2-dt^2)\eqno (13)
$$

where

$$
h_i=\frac d{|{\bf r-r}_i|}\eqno (14)
$$

\section{The Computer Program}

I constructed a symbolic computer program to help find this solution. It was
a bit like using a canon to shoot down a pea. It was written in both
Mathematica and MapleV.

The program takes as input, the coordinates, the covariant components of the
metric and the covariant components of the electromagnetic vector potential.
It then outputs the determinant of the metric, the contravariant components
of the metric, the electromagnetic field strengths, the current density, the
Maxwell stress-energy tensor, the Christoffel
symbols, the curvature tensor, the Ricci tensor, the scalar curvature, the
Einstein tensor and the Einstein equations of motion. If the last is not
satisfied, it gives the remaining terms by which it fails. I will restrict
the following detailed discussion to MapleV.

In order to distinguish between the covariant and the contravariant
components of the metric, I used gl[i,j] for the covariant and gu[i,j] for
the contravariant components. The program itself very closely follows the
equations one would normally write down while doing a calculation. As an
example I give the lines for computing the Christoffel symbols.\\

gam:=array(1..4,1..4,1..4):

for i to 4 do for j to 4 do for k to 4 do

gam[i,j,k]:=1/2*simplify( sum( 'gu[i,l]*(diff( gl[l,j], x[k]) + diff(
gl[l,k], x[j])

- diff( gl[j,k], x[l]))','l'=1..4)) od: od: od: \vspace{12pt} \hfill (15)

The input is accomplished by typing the data into a first program rather
than interacting with the computer directly. This is then called by a second
program. This system enables the input program to be short enough so that
one can save a number of them efficiently. Results one wishes to save are
specified in the second program and may be used as input for further work.

\section{Summary}

In sections 2 and 3, we have two very different solutions to Eqs.(1) and (2)
for a point charge. Eq.(11) is the well known solution which asymptotically
approaches flat space while for Eq.(5) $ds^2\rightarrow 0$ asymptotically.
In section 4 we then discussed how concentration of the metric about a
charge to within distances tiny compared to atomic sizes enables one to
obtain an approximate solution for the case of many charges separated by
distances of atomic dimensions.

Of course, quantum mechanical field theoretic considerations will play a
dominant role at the small distances discussed in this paper.
However, a model such as the one presented here might
serve as the background field in a quantization procedure.

While the computer program is not a substitute for keen insight, nor even
pen and paper, it should be helpful in looking at more complex solutions to
the Einstein-Maxwell problem.

\newcommand{\RMP}[3]{{\em Rev. Mod. Phys.} {\bf #1}, #2 (19#3)}
\newcommand{\Rep}[3]{{\em Phys. Rep.} {\bf #1}, #2 (19#3)}
\newcommand{\Ann}[3]{{\em Annals of Phys.} {\bf #1}, #2 (19#3)}
\newcommand{\NS}[3]{{\em Nucl. Sci.} {\bf #1}, #2 (19#3)}
\newcommand{\PR}[3]{{\em Phys. Rev.} {\bf #1}, #2 (19#3)}
\newcommand{\PRL}[3]{{\em Phys. Rev. Letts.} {\bf #1}, #2 (19#3)}
\newcommand{\PL}[3]{{\em Phys. Letts.} {\bf #1}, #2 (19#3)}
\newcommand{\NPB}[3]{{\em Nucl. Phys. B} {\bf #1}, #2 (19#3)}


\begin{thebibliography}{9}
\bibitem{4}  We note that if $e=4.8\times 10^{-10}$ esu (the electron
charge), then $d=1.4\times 10^{-34}$ cm.
\end{thebibliography}
\end{document}